\documentclass[a4paper,11pt]{article}

\usepackage{jheppub} 

\usepackage[T1]{fontenc} 

\usepackage{graphicx} 
\usepackage{amsthm}
\usepackage{tikz-network}
\usepackage{subfloat}
\usepackage{subfig}

\usepackage{bbm}

\newtheorem{thm}{Theorem}[section]
\newtheorem{prop}[thm]{Proposition}
\newtheorem{defn}[thm]{Definition}

\newcommand{\abs}[1]{\left| #1 \right|} 
\newcommand{\ket}[1]{| #1 \rangle } 
\newcommand{\braket}[2]{\langle  #1 \vphantom{#2} | #2 \vphantom{#1} \rangle } 
 
\newcommand{\diracprod}[3]{\left\langle  #1 \vphantom{#2#3} \right|#2 \left| #3 \vphantom{#1#2} \right\rangle } 
\let\baraccent=\= 
\renewcommand{\=}[1]{\stackrel{#1}{=}} 

\def\A{\mathcal{A}}
\def\B{\mathcal{B}}
\def\C{\mathcal{C}}

\def\N{\mathcal{N}}

\def\O{\mathcal{O}}
\def\P{\mathcal{P}}

\def\L{\mathcal{L}}

\def\H{\mathcal{H}}
\def\R{\mathcal{R}}

\def\Tr{\text{Tr}}

\def\Srel{S_{\text{rel}}}
\def\id{\mathbbm{1}}

\def\AIR{\mathcal{A}_{I\cup R}}
\def\AX{\mathcal{A}_{X}}
\def\ARAD{\mathcal{A}_{\text{Rad}}}
\def\ABH{\mathcal{A}_{\text{BH}}}

\def\HIR{\mathcal{H}^{sc}_{I\cup R}}
\def\HX{\mathcal{H}^{sc}_{X}}
\def\HRAD{\mathcal{H}^{qg}_{\text{Rad}}}
\def\HBH{\mathcal{H}^{qg}_{\text{BH}}}

\def\TPsi{\widetilde{\Psi}}
\def\TPhi{\widetilde{\Phi}}

\title{An algebraic description of the Page transition}

\author{Haocheng Zhong$^{a}$}

\affiliation{$^a$Shing-Tung Yau Center and School of Physics, Southeast University, Nanjing 210096, China}

\emailAdd{zhonghaocheng@seu.edu.cn}

\abstract{In this work, we develop an algebraic description of the Page transition, a key feature in black hole evaporation where the entropy of Hawking radiation follows a unitary Page curve instead of monotonically increasing. By applying concepts from approximate quantum error correction with complementary recovery, we characterize the Page transition as a phase transition in channel recovery. We then generalize the description to infinite-dimensional settings using algebraic relative entropy, which remains valid even in type III factors. For type I/II factors, explicit probes based on relative entropy differences are derived, serving as indicators for the transition at the Page time. }

\begin{document} 
\maketitle
\flushbottom

\section{Introduction}


The Page curve \cite{Page:1993wv,Page:2013dx} is a theoretical prediction for how the fine-grained entropy of Hawking radiation should behave over the lifetime of an evaporating black hole if the evaporation process is unitary (i.e., if information is conserved), which is central to modern discussions of the black hole information paradox \cite{Mathur:2009hf,Almheiri:2012rt,Penington:2019npb,Almheiri:2020cfm}. The paradox originates from Hawking's calculations \cite{Hawking:1975vcx,Hawking:1976ra} in 1970s which suggested that Hawking radiation is purely thermal and random. Therefore, the entropy of radiation should monotonically increase throughout the black hole evaporation. If this were true, when the black hole vanishes, the information about what fell into it would be irretrievably lost, violating the quantum mechanical principle of unitarity. On the other hand, the Page cure suggested that the entropy should reach a maximum at a point known as the Page time, after which the entropy monotonically decreases and returns to zero. Recently, the island formula \cite{Penington:2019npb,Almheiri:2019hni,Penington:2019kki,Almheiri:2019qdq,Almheiri:2019psf} is proposed to resolve the paradox, which provides a method for calculating the fine-grained entropy of Hawking radiation in a way that preserves unitarity, resulting in the Page curve rather than the information-loss-inducing curve predicted by Hawking's original calculation.

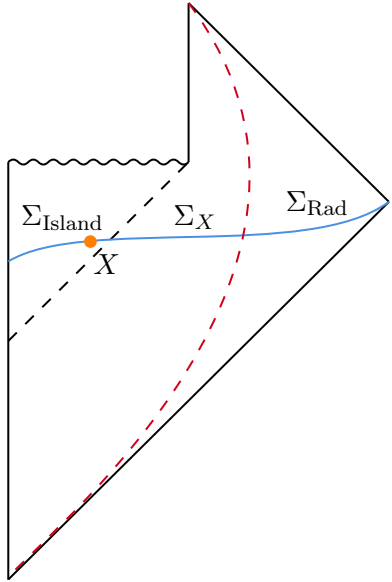
\begin{figure}[h]
	\centering
	\tikzset{every picture/.style={line width=0.75pt}} 
	\begin{tikzpicture}[x=0.75pt,y=0.75pt,yscale=-1,xscale=1]
		
		\draw    (210,90) -- (210,300) ;
		\draw    (210,90) .. controls (211.67,88.33) and (213.33,88.33) .. (215,90) .. controls (216.67,91.67) and (218.33,91.67) .. (220,90) .. controls (221.67,88.33) and (223.33,88.33) .. (225,90) .. controls (226.67,91.67) and (228.33,91.67) .. (230,90) .. controls (231.67,88.33) and (233.33,88.33) .. (235,90) .. controls (236.67,91.67) and (238.33,91.67) .. (240,90) .. controls (241.67,88.33) and (243.33,88.33) .. (245,90) .. controls (246.67,91.67) and (248.33,91.67) .. (250,90) .. controls (251.67,88.33) and (253.33,88.33) .. (255,90) .. controls (256.67,91.67) and (258.33,91.67) .. (260,90) .. controls (261.67,88.33) and (263.33,88.33) .. (265,90) .. controls (266.67,91.67) and (268.33,91.67) .. (270,90) .. controls (271.67,88.33) and (273.33,88.33) .. (275,90) .. controls (276.67,91.67) and (278.33,91.67) .. (280,90) .. controls (281.67,88.33) and (283.33,88.33) .. (285,90) .. controls (286.67,91.67) and (288.33,91.67) .. (290,90) .. controls (291.67,88.33) and (293.33,88.33) .. (295,90) .. controls (296.67,91.67) and (298.33,91.67) .. (300,90) -- (300,90) ;
		\draw    (300,90) -- (300,10) ;
		\draw    (300,10) -- (400,110) ;
		\draw    (400,110) -- (210,300) ;
		\draw  [dash pattern={on 4.5pt off 4.5pt}]  (210,180) -- (300,90) ;
		\draw [color={rgb, 255:red, 74; green, 144; blue, 226 }  ,draw opacity=1 ]   (210,140) .. controls (247.2,117) and (360,140) .. (400,110) ;
		\draw [color={rgb, 255:red, 208; green, 2; blue, 27 }  ,draw opacity=1 ] [dash pattern={on 4.5pt off 4.5pt}]  (300,10) .. controls (404.2,149.27) and (207.2,293) .. (210,300) ;
		
		\draw (215,110.4) node [anchor=north west][inner sep=0.75pt]    {$\Sigma_{\text{Island}}$};
		\draw (291,110) node [anchor=north west][inner sep=0.75pt]    {$\Sigma_{X}$};
		\draw (347,103) node [anchor=north west][inner sep=0.75pt]    {$\Sigma_{\text{Rad}}$};
		\draw (251,135) node [anchor=north west][inner sep=0.75pt]    {$X$};
		
		\filldraw[orange]  (251,130) circle (2pt);
	\end{tikzpicture}
	\caption{The Penrose diagram of an evaporating black hole. A Cauchy slice (blue solid curve) is divided into three regions $\Sigma_{\text{Island}},\Sigma_{X},\Sigma_{\text{Rad}}$ by a codimension-2 surface $X$ (orange dot) and a cutoff surface (red dotted line).}
	\label{fig:BH}
\end{figure}

To illustrate, the Penrose diagram of a black hole during evaporation is depicted in Fig.\ref{fig:BH}, where a Cauchy slice is factorized into three spacelike regions: (a) an interior ``island'' enclosed by a codimension-2 surface $X$, which we denote as $\Sigma_{\text{Island}}$; (b) the region bounded by $X$ and a cutoff surface near the asymptotic boundary, which is denoted as $\Sigma_X$; (c) the region between the cutoff surface and the spacelike infinity, and we denote it as $\Sigma_{\text{Rad}}$ where external observers collects the black hole radiation. The entropy of the black hole for this Cauchy slice can be characterized by the following formula, 
\begin{equation}\label{eq:island1}
	S_{\text{BH}}=\text{min}_X\left\{\operatorname{ext}_X\left[\frac{A_X}{4 G}+S_{\text {semi-cl }}\left(\Sigma_X\right)\right]\right\},
\end{equation}
where $A_X$ is the area of the entangling surface $X$, and $S_{\text {semi-cl }}\left(\Sigma_X\right)$ is the von Neumann entropy of the quantum states on $\Sigma_X$ in the semi-classical description, by which we mean that the entropy is calculated by the standard methods of quantum field theory in curved spacetime. Meanwhile, $S_{\text{BH}}$ should be regarded as the von Neumann entropy in a complete quantum gravity description, which is yet to be well established. The condition ``$\operatorname{ext}_X$'' selects the saddle points of a generalized entropy insides the square bracket among all possible configurations of $X$, and if there exists several options, we choose the minimal one.

\begin{figure}[h]
	\centering

	\tikzset{every picture/.style={line width=0.75pt}} 
	
	\begin{tikzpicture}[x=0.75pt,y=0.75pt,yscale=-1,xscale=1]
		
		\draw    (150,270) -- (508,270) ;
		\draw [shift={(510,270)}, rotate = 180] [color={rgb, 255:red, 0; green, 0; blue, 0 }  ][line width=0.75]    (10.93,-3.29) .. controls (6.95,-1.4) and (3.31,-0.3) .. (0,0) .. controls (3.31,0.3) and (6.95,1.4) .. (10.93,3.29)   ;
		\draw    (150,270) -- (150,72) ;
		\draw [shift={(150,70)}, rotate = 90] [color={rgb, 255:red, 0; green, 0; blue, 0 }  ][line width=0.75]    (10.93,-3.29) .. controls (6.95,-1.4) and (3.31,-0.3) .. (0,0) .. controls (3.31,0.3) and (6.95,1.4) .. (10.93,3.29)   ;
		\draw [color={rgb, 255:red, 74; green, 144; blue, 226 }  ,draw opacity=1 ]   (470,90) -- (150,270) ;
		\draw [color={rgb, 255:red, 245; green, 166; blue, 35 }  ,draw opacity=1 ]   (150,90) -- (470,270) ;
		\draw [color={rgb, 255:red, 208; green, 2; blue, 27 }  ,draw opacity=1 ]   (160,270) -- (310,190) ;
		\draw [color={rgb, 255:red, 208; green, 2; blue, 27 }  ,draw opacity=1 ]   (450,270) -- (310,190) ;
		\draw  [dash pattern={on 4.5pt off 4.5pt}]  (310,170) -- (310,270) ;
		
		\draw (527,272.4) node [anchor=north west][inner sep=0.75pt]    {$t$};
		\draw (307,272.4) node [anchor=north west][inner sep=0.75pt]    {$t_{P}$};
		\draw (241,112.4) node [anchor=north west][inner sep=0.75pt]    {$S_{\text{Rad}}^{\text{island}}$};
		\draw (350,112.4) node [anchor=north west][inner sep=0.75pt]    {$S_{\text{Rad}}^{\text{no-island}}$};
		\draw (369,242.4) node [anchor=north west][inner sep=0.75pt]    {$S_{\text{Rad}}$};

	\end{tikzpicture}
	
	\caption{A sketch of the Page curve (red line) for the entropy of Hawking radiation, which is upper bounded by the two saddle-point solutions: the vanishing island solution $S_{\text{Rad}}^{\text{no-island}}$ (blue line) and the non-vanishing solution $S_{\text{Rad}}^{\text{island}}$ (yellow line). The transition of the two dominant solutions happens at the Page time $t_P$, which is call the Page transition.}
	\label{fig:Page}
\end{figure}
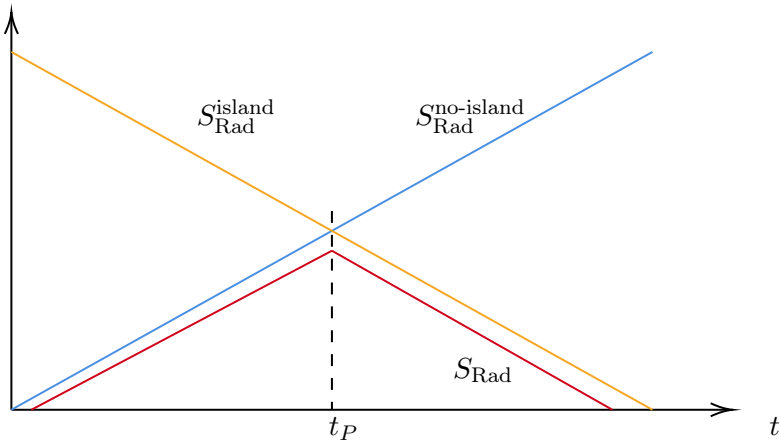

When the black hole is formed from an initially pure state, we expect that the evaporating black hole itself and the radiation already emitted during the evaporation make a pure state, such that
the entropy of Hawking radiation should be equal to the entropy of the black hole, giving a similar island formula,
\begin{equation}\label{eq:island2}
	S_{\text {Rad }}=\text{min}_X\left\{\operatorname{ext}_X\left[\frac{A_X}{4 G}+S_{\text {semi-cl }}\left(\Sigma_{\text {Rad }} \cup \Sigma_{\text {Island }}\right)\right]\right\},
\end{equation}
where we have $S_{\text {semi-cl }}\left(\Sigma_{\text {Rad }} \cup \Sigma_{\text {Island }}\right)=S_{\text {semi-cl }}\left(\Sigma_X\right)$ due to the fact that $\Sigma_{\text {Rad }} \cup \Sigma_{\text {Island }}\cup \Sigma_X$ prepares a pure state over the Cauchy slice. The formula \eqref{eq:island2} solves the puzzle that the fine-grained entropy of black hole radiation should satisfy the Page curve, which states that the entropy should monotonically increase before the Page time $t_P$, and monotonically decrease after $t_P$, as depicted in Fig.\ref{fig:Page}. The main feature of \eqref{eq:island2} is that, the entropy of radiation should be regarded as a competition of two saddle-point solutions of the extremal condition in the gravitational path integral description \cite{Penington:2019npb,Penington:2019kki,Almheiri:2020cfm,Almheiri:2019qdq}: the solution with the island vanishing (so as the boundary $X$), and the solution with a non-vanishing island bounded by a non-vanishing $X$, i.e.
\begin{equation}
	S_{\text{Rad}}=\min\left\{S_{\text{Rad}}^{\text{island}},S_{\text{Rad}}^{\text{no-island}}\right\},
\end{equation}
where the two terms on the RHS behave differently, see also Fig.\ref{fig:Page}. The solution $S_{\text{Rad}}^{\text{no-island}}$ with vanishing island, which coincides with Hawking's calculation, monotonically increases during the whole evaporation process. Meanwhile, the solution $S_{\text{Rad}}^{\text{island}}$ with non-vanishing island, which captures the thermodynamic properties of the black hole, monotonically decreases. The two solutions are course-grained descriptions for the entropy, such that the fine-grained entropy should be upper-bounded by the two, hence giving a phase transition called the Page transition at the Page time $t_P$.

It was recently shown that \cite{Zhong:2024fmn} the Page transition can be regarded as a phase transition in approximate quantum error correction (AQEC). To be more specific, the island formula \eqref{eq:island2} implies that the external observers who collects the black hole radiation detects the emergence of an island region at the Page time. The island region is the black hole interior ``accessible'' to the external observers, which is related to the notion of reconstruction insides the island. Such reconstruction can be well described by a series of theorems via quantum error correction \cite{Almheiri:2014lwa,Dong:2016eik,Harlow:2016vwg,Kamal:2019skn,Kang:2018xqy,Xu:2024xbz,Crann:2024gkv}. We call them the reconstruction theorems, which were originally proved for explaining the problem of bulk reconstruction in the AdS/CFT correspondence \cite{Maldacena:1997re,Witten:1998qj,Gubser:1998bc}. See also \cite{Kang:2019dfi,Faulkner:2020hzi,Gesteau:2020hoz,Gesteau:2020rtg,Akers:2021fut,Gesteau:2023hbq,Crann:2024gkv,Gomez:2024fij,Gomez:2024cjm,Engelhardt:2023xer,vanderHeijden:2024tdk} for an incomplete list of related developments. In \cite{Zhong:2024fmn}, the author applied the following version of the reconstruction theorem to the black hole evaporation,
\begin{thm}\label{thm:RecThm}
	Given a finite-dimensional Hilbert space $\H$ (called the physical space) which is decomposed into two parts $\H=\H_{A}\otimes\H_{\bar{A}}$ and a code space $\H_{code}$ with condition $dim\H_{code}\leq\abs{A}$, the isometric encoding $V: \H_{code}\rightarrow \H$ induces a quantum channel:
	\begin{align}
		\N: ~&s(\H_{code})\rightarrow s(\H_A),\quad\tilde{\rho}\mapsto \rho_A\equiv\Tr_{\bar{A}}\left(V\tilde{\rho}V^\dagger\right),
	\end{align}
	where $s(\H_{(\cdot)})$ denotes the set of all density operators in $\H_{(\cdot)}$. Then the following statements are equivalent:
	\begin{enumerate}

		\item Given $\tilde{\rho}\in s(\H_{code})$, we have
		\begin{align}
			&S(\rho_A)=\mathcal{L}_A+S(\tilde{\rho})\label{rt1}
		\end{align}
		where $\mathcal{L}_A$ is some constant.
		
		\item
		
		Given two density matrices $\tilde{\rho},\tilde{\sigma}\in s(\H_{code})$, we have
		\begin{align}
			\Srel(\rho_A|\sigma_A)&=\Srel(\tilde{\rho}|\tilde{\sigma}).
		\end{align}

		\item For any operator $\tilde{O}\in\mathcal{L}(\H_{code})$, there exists an operator $O_A\in\mathcal{L}(\H_A)$ such that
		\begin{equation}
			O_AV\ket{\tilde{\psi}}=V\tilde{O}\ket{\tilde{\psi}},\quad \forall \ket{\tilde{\psi}}\in \H_{code}.
		\end{equation}

	\end{enumerate}

\end{thm}
It has been argued that the quantum states on the island and the radiation region together form a code space in semi-classical description, and the radiation collected by the external observers forms a physical space in the complete quantum gravity description. Therefore, after identifying the statement 1 of the theorem \ref{thm:RecThm} with the formula of entropy of radiation \eqref{eq:island2}, the statement 3 states that one is able to manipulate the degrees of freedoms in the island by manipulating the the degrees of freedoms in the radiation only. In this sense, the channel $\N$ represents some kind of a tunnel for information transferring between the island and the exterior of the black hole. However, such channel $\N$ can only appear after the appearance of a non-vanishing island, so that the Page transition can also be regarded as the emergence of the channel $\N$ at the Page time. Then one may wonder whether we can have a probe for the emergence of the channel $\N$ to describe the Page transition. Interestingly, the statement 2 of the theorem \ref{thm:RecThm}, which involves the equality between two relative entropies, is closely related to the exactness of recovery for the channel. The related theorems are collectively shown as follows \cite{ohya2004quantum},
\begin{thm}\label{thm:relentropymono}
	Let $\N$ be a quantum channel from system $A$ to system $B$, then we have
	\begin{equation}
		\text{(Monotonicity)}\quad \Srel(\rho|\sigma)\geq \Srel(\N(\rho)|\N(\sigma)),\quad \forall \rho,\sigma\in s(\H_A),
	\end{equation}
	and the inequality is saturated if and only if $\N$ is exactly reversible, i.e. there exists another channel $\R$ (called the recovery channel) such that
	\begin{equation}
		\text{(Exact recovery)}\quad \tilde{\rho}\equiv(\mathcal{\R} \circ \mathcal{N})(\rho) = \rho,\quad \forall\rho\in s(\H_{code}).
	\end{equation}
\end{thm}
In the language of quantum information, the channel during black hole evaporation is not exactly reversible before the Page time, but then becomes exactly reversible after the Page time. The intermediate period, which is the Page transition, should correspond to a ``phase transition'' of the recovery for the channel, and the channel should be approximately reversible during the period. The theorem \ref{thm:relentropymono} then helps to give a criteria that for some infinitesimal value $\delta$ which is related to the properties of the black hole, we have
\begin{equation}\label{eq:probe}
	\Srel(\tilde{\rho}_{I\cup R}(t)|\tilde{\sigma}_{I\cup R}(t))-\Srel(\rho_{\text{Rad}}(t)|\sigma_{\text{Rad}}(t))\begin{cases}
		>\delta,  & \text{if $t<t_{P}$} \\
		=\delta, & \text{if $t=t_{P}$}\\
		\leq\delta, & \text{if $t>t_{P}$}
	\end{cases}
\end{equation}
which works as a probe for the Page transition.

In fact, whenever the channel between the island and the radiation exists, there exists a dual channel which relates the region $\Sigma_X$ to the black hole itself, which is due to the existence of the parallel formula \eqref{eq:island1} of the entropy of black hole. As we will show in the paper, the two channels are associated to an error-correcting code with complementary recovery. For the reconstruction theorem to be applicable in this case, we will apply an enhanced version of theorem \ref{thm:RecThm}, which we refer to as the reconstruction theorem with complementary recovery. Then following the same arguments, there also exists a similar probe as \eqref{eq:probe} for the dual channel to experience the phase transition during the Page transition.

Note the above arguments are only limited in the cases of finite-dimensional black holes, as we \emph{a priori} assume that the central dogma \cite{Almheiri:2020cfm,Strominger:1996sh,Dabholkar:2014ema} to hold in general. However, the reconstruction theorems themselves admit algebraic generalizations \cite{Kamal:2019skn,Kang:2018xqy,Xu:2024xbz,Crann:2024gkv} which are capable of giving algebraic descriptions for infinite-dimensional Hilbert spaces, it is then plausible to generalize the above arguments for the infinite-dimensional evaporation black holes, which is one of the main tasks in the paper. We will give explicit calculations to show that, the algebraic versions of probes similar to \eqref{eq:probe} can be directly derived using the algebraic versions of the island formulae \eqref{eq:island1} and \eqref{eq:island2}. Also we would like to point out that another interesting feature of \eqref{eq:probe} is that the relative entropy is well-defined for von Neumann algebras of general types, including algebras of type III where the von Neumann entropies in either \eqref{eq:island1} or \eqref{eq:island2} are ill-defined. In this sense, the probes using relative entropy are more general, and we will give more comments on these issues in the discussions.

The work is organized as follows. In section \ref{sec:pre}, we introduce the reconstruction theorems, and some basics of von Neumann algebras with algebraic generalizations for entropies. In section \ref{sec:complementary recovery}, we apply the reconstruction theorem with complementary recovery to the black hole evaporation. In section \ref{sec:algebraic description}, we discuss how to generalize our arguments into algebraic description in factors of type I/II, and give a formal proof. We end with section \ref{sec:discussions} where we give some comments about related works.

\section{Preliminary}\label{sec:pre}

\subsection{The reconstruction theorem with complementary recovery}

The finite-dimensional reconstruction theorem based on an error-correcting code with complementary recovery is stated as follows \cite{Harlow:2016vwg},

\begin{thm}\label{thm:RecThm2}
	Given a finite-dimensional system called the physical space $\H_{phys}=\H_{A}\otimes\H_{\bar{A}}$ and a code space $\H_{code}=\H_{a}\otimes\H_{\bar{a}}$ with conditions $\abs{a}\leq\abs{A},\abs{\bar{a}}\leq\abs{\bar{A}}$, the isometry\footnote{An isometry $V: \H_A\rightarrow \H_B~(\abs{A}\leq \abs{B})$ satisfies $V^\dagger V=I_A$ and $VV^\dagger=\Pi_{B}$ which is a projector onto $V(\H_A)\subset\H_B$.} $V: \H_{code}\rightarrow \H_{phys}$ induces two quantum channels:
	\begin{align*}
		\N: ~\tilde{\rho}_a\mapsto \rho_A\equiv\Tr_{\bar{A}}\left(V\tilde{\rho}_aV^\dagger\right),\quad \N': ~\tilde{\rho}_{\bar{a}}\mapsto \rho_{\bar{A}}\equiv\Tr_{A}\left(V\tilde{\rho}_{\bar{a}}V^\dagger\right),
	\end{align*}
	
	then the following statements are equivalent:
	
	\begin{enumerate}
		
		\item Given $\widetilde{\O}_a$ on $\H_a$: $\exists \O_A\in\mathcal{L}(\H_A)\Rightarrow \O_AV|\tilde{\phi}\rangle=V\widetilde{\O}_a|\tilde{\phi}\rangle$; likewise for $\widetilde{\O}_{\bar{a}}$.
		
		\item Given $\tilde{\rho},\tilde{\sigma}$ on $\H_{code}$: $\Srel(\rho_A|\sigma_A)=\Srel(\tilde{\rho}_a|\tilde{\sigma}_a),~\Srel(\rho_{\bar{A}}|\sigma_{\bar{A}})=\Srel(\tilde{\rho}_{\bar{a}}|\tilde{\sigma}_{\bar{a}})$.
		
		\item 
		
		Given $\widetilde{\O}_{a}$ on $\H_a$: $[\widetilde{\O}_{a},V^\dagger \P _{\bar{A}}V]=0,\forall \P _{\bar{A}}\in\mathcal{L}(\H_{\bar{A}})$; likewise for $\widetilde{\O}_{\bar{a}}$.
		
		\item Given $\tilde{\rho}$ on $\H_{code}$: $S(\rho_A)=\mathcal{L}_A+S(\tilde{\rho}_a),~S(\rho_{\bar{A}})=\mathcal{L}_{\bar{A}}+S(\tilde{\rho}_{\bar{a}})$ where $\mathcal{L}_A=\mathcal{L}_{\bar{A}}$ are some constants.

	\end{enumerate}
	
\end{thm}

The theorem can be proved without referring to any physical interpretation, see for example appendix A of \cite{Zhong:2024fmn}. The basic idea of the theorem states that whenever any one of the above statements holds, the degrees of freedom in $\H_{a},\H_{\bar{a}}$ can be reconstructed from $\H_{A},\H_{\bar{A}}$ respectively. By ``reconstruction'' of degrees of freedom we are actually referring to the statement 1 where there also exists a dual operator in the reconstructed space for an arbitrary operator in the code subspace, hence giving the name the ``reconstruction theorem''. Compared to theorem \ref{thm:RecThm}, the reconstructions hold pair-wisely for $\H_{A}$ and $\H_{\bar{A}}$, which we call the complementary recovery. In the language of quantum information, the statement 1 is also related to the notion of quantum error correction, in the sense that information in the code space is protected by enlarging the space (i.e. an isometry into the physical space) against the ``erasure''  (i.e. some extra degrees of freedom in the physical space are inaccessible to the operators acting on the code space). For more comments about the theorem and the quantum error correction, see \cite{Harlow:2018fse}.

The above statements are more comprehensive in some concrete examples, e.g. the AdS/CFT correspondence. In this case, the four statements respectively corresponds to (1) entanglement wedge reconstruction (or subregion duality) \cite{Czech:2012bh,Hamilton:2006az,Morrison:2014jha,Bousso:2012sj,Bousso:2012mh,Hubeny:2012wa,Wall:2012uf,Headrick:2014cta,Jafferis:2015del,Dong:2016eik}; (2) the Jafferis-Lewkowycz-Maldacena-Suh (JLMS) formula \cite{Jafferis:2015del}; (3) radial commutativity \cite{Polchinski:1999yd,Almheiri_2015,Harlow:2018fse}; (4) the Ryu-Takayanagi (RT) formula \cite{Ryu:2006bv,Hubeny:2007xt,Casini_2011,Lewkowycz:2013nqa,Nishioka:2018khk,Faulkner:2013ana,Engelhardt:2014gca}. While in this work, we will apply the theorem \ref{thm:RecThm2} to the black hole evaporation by examining whether the prerequisites of the theorem are satisfied. More explicitly, we will relate the statement 4 to the formulae \eqref{eq:island1} and \eqref{eq:island2}, then argue that the statement 2 implies a probe for the Page transition.

\subsection{Basics of von Neumann algebra}\label{subsec:vN alg}

In this subsection, we introduce some basics of von Neumann algebra. The theorems and the statements listed here are demonstrated without proofs, one can consult \cite{Vaughan2009,Sorce:2023fdx,reed1972methods,takesaki2006tomita} for more rigorous arguments, and also \cite{Witten:2018zxz,Kang:2018xqy,Kang:2019dfi,Xu:2024xbz} for relevant contents.

\begin{defn}
	A linear operator on a Hilbert space $\H$ is a linear map from (a subspace of) $\H$ into $\H$. The set of all such operators is denoted by $\L(\H)$.
\end{defn}

\begin{defn}
	A bounded operator is a linear operator $\O$ which satisfies $||\O\ket{\psi}||\leq K||\ket{\psi}||,~\forall\ket{\psi}\in\H$ for some $K\in\mathbb{R}$. The infimum of all such $K$ is called the norm of $\O$. The algebra of all bounded operators on $\H$ is denoted by $\B(\H)\subset \L(\H)$.
\end{defn}

\begin{defn}
	The commutant of a subset $S\subset \B(\H)$ is a subset $S'\subset \B(\H)$ defined by
	\begin{equation}
		S'\equiv\{\O\in\B(\H)|[\O,\P]=0,~\forall \P\in S\}
	\end{equation}
	i.e. every element in $S'$ commutes with all elements in $S$.
\end{defn}

\begin{defn}
	The hermitian conjugate (or adjoint) of an operator $\O$ is an operator $\O^\dagger$ satisfying $\braket{\psi}{\O\xi}=\braket{\O^\dagger \psi}{\xi}$. A hermitian (or self-adjoint) operator $\O$ satisfies $\O=\O^\dagger$.
\end{defn}

\begin{defn}
	A von Neumann algebra on $\H$ is a subalgebra $\A\subset\B(\H)$ satisfying
	\begin{itemize}
		\item $\id\in \A$,
		
		\item $\A$ is closed under hermitian conjugation,
		
		\item $\A''=\A$.
		
	\end{itemize} 
\end{defn}

\begin{defn}
	A von Neumann algebra $\A$ is a factor if it has a trivial center $\mathcal{Z}$:
	\begin{equation}
		\mathcal{Z}\equiv \A\cap\A'=\{\lambda \id|\lambda\in\C\}
	\end{equation}
	otherwise $\A$ is called a non-factor.
\end{defn}

A non-factor can always be ``decomposed'' into factors \cite{Sorce:2023fdx,Harlow:2016vwg} such that whenever we consider classifications of von Neumann algebra, we only need to consider factors, which are already classified into three types: type I/II/III. For factors of each types, some notions may be well-defined or not, as summarized as follows,

$$
\begin{array}{|c|c|c|c|c|c|}
	\hline {\text {Type}} & \H=\H_A \otimes \H_B & \Tr & \rho_\psi & S(\psi;\A)  & S_{\text {rel }}(\psi|\xi;\A) \\
	\hline \text { I } & \checkmark & \checkmark & \checkmark & \checkmark  & \checkmark \\
	\hline \text { II } & \times & \checkmark & \checkmark & \checkmark &  \checkmark \\
	\hline \text { III } & \times & \times & \times & \times  & \checkmark \\
	\hline
\end{array}
$$
where we write $\H=\H_A \otimes \H_B$ to denote that the Hilbert space can be decomposed according to subregions. Note the relative entropy has an algebraic generalization denoted as $S_{\text {rel }}(\psi|\xi;\A)$ for factors of any type, while the algebraic von Neumann entropy denoted as $S(\psi;\A)$ can only be well-defined in type I/II due to the lack of well-defined notions of the trace function and density operators in factors of type III. Before we introduce the algebraic definitions of these entropies, we need to learn some backgrounds about the Tomita-Takesaki theory (also called the modular theory).

\begin{defn}
	A subset $\mathcal{H}_{0}\subset\mathcal{H}$ is dense in $\mathcal{H}$ if for every vector $\ket{\psi} \in \mathcal{H}$ and for every $\epsilon>0$, there exists a vector $\ket{\phi} \in \mathcal{H}_{0}$ such that $\|\ket{\psi}-\ket{\phi}\|<\epsilon$.
\end{defn}

\begin{defn}
	$\ket{\psi}\in\H$ is cyclic with respect to a von Neumann algebra $\A$ if $\A\ket{\psi}\equiv\{\O\ket{\psi}|\forall \O\in \A\}$ is dense in $\H$.
\end{defn}


\begin{defn}
	$\ket{\psi}\in\H$ is separating with respect to a von Neumann algebra $\A$ if $\O\ket{\psi}=0 $ implies $\O=0$ for $\O\in\A$.
\end{defn}


\begin{thm}\label{thm:cs}
	$\ket{\psi}\in\H$ is separating with respect to $\A$ if and only if $\ket{\psi}\in\H$ is cyclic with respect to $\A'$, and vice versa.
\end{thm}

A direct consequence is that when we assume $\ket{\psi}\in\H$ is both cyclic and separating with respect to $\A$, then $\ket{\psi}\in\H$ is also both cyclic and separating with respect to $\A'$.

\begin{defn}
	A relative Tomita operator on $\A$ is an anti-linear operator satisfying\footnote{Some literatures use the convention that interchanges the positions of $\psi$ and $\xi$ in the subscript of $S$.}
	\begin{equation}
		S_{\xi|\psi}\left(\O\ket{\psi}\right)=\O^\dagger \ket{\xi},\quad \forall \O\in\A.
	\end{equation}
\end{defn}

Notice that $S_{\xi|\psi}$ is densely defined, i.e. whose domain is a dense subset of $\H$ if and only if $\ket{\psi}$ is cyclic and separating with respect to $\A$. The cyclic condition ensures that the domain $\O\ket{\psi}$ is dense while the separating condition is to avoid the possibility that $\O\ket{\psi}=0,~ \O^\dagger\ket{\xi}\neq 0$. Hereafter we mostly assume the cyclic separating condition of $\ket{\psi}$ for $S_{\xi|\psi}$.

\begin{prop}
	Provided both $\ket{\psi},\ket{\xi}$ are cyclic and separating with respect to $\A$, we have
	\begin{equation}
		S_{\xi|\psi}^{-1}=S_{\psi|\xi},\quad S^\dagger_{\xi|\psi}=S'_{\xi|\psi},
	\end{equation}
	where $S'_{\xi|\psi}$ is a relative Tomita operator on $\A'$.
\end{prop}

\begin{defn}
	Provided $\ket{\psi}$ is cyclic and separating with respect to $\A$, the relative modular operator on $\A$ is defined by\footnote{There is an equivalent definition of the relative modular operator via the unique polar decomposition $S_{\xi|\psi}=J_{\xi|\psi}\Delta_{\xi|\psi}^{\frac{1}{2}}$ with $J_{\xi|\psi}$ anti-unitary.}
	\begin{equation}
		\Delta_{\xi|\psi}\equiv S^\dagger_{\xi|\psi}S_{\xi|\psi},
	\end{equation}
	and the relative modular Hamiltonian on $\A$ is defined by 
	\begin{equation}
		h_{\xi|\psi}\equiv-\log \Delta_{\xi|\psi}.
	\end{equation}
\end{defn}

\begin{prop}
	The relative modular operator $\Delta_{\xi|\psi}$ on $\A$ is Hermitian.
\end{prop}

\begin{prop}
	Provided both $\ket{\psi},\ket{\xi}$ are cyclic and separating with respect to $\A$, we have
	\begin{equation}
		\Delta_{\psi|\xi}^{-1}=\Delta'_{\xi|\psi}\quad\Rightarrow\quad h_{\psi|\xi}=-h_{\xi|\psi}'
	\end{equation}
	where $\Delta'_{\xi|\psi}$ and $h_{\xi|\psi}'$ are the relative modular operator and the relative modular Hamiltonian on $\A'$ respectively.
\end{prop}

\begin{thm}[Splittable condition]\label{thm:split}
	
	In factors of type I/II, we have
	\begin{equation}\label{eq:Hamiltonian splittable infin}
		\Delta_{\phi|\psi}=\rho_{\phi;\A}\otimes\rho_{\psi;\A'}^{-1}.
	\end{equation}
\end{thm}
Physically, \eqref{eq:Hamiltonian splittable infin} implies that the usual modular Hamiltonian of a physical system is splittable. The condition is crucial in deriving the generalized second law for the generalized entropy in gravitational background, see \cite{Jensen:2023yxy,Faulkner:2024gst}.

\begin{defn}
	Provided $\ket{\psi}$ is cyclic and separating with respect to $\A$, we define the Tomita operator $S_{\psi}\equiv S_{\psi|\psi}:\O\ket{\psi}\mapsto\O^\dagger \ket{\psi}$; the modular operator $\Delta_{\psi}\equiv\Delta_{\psi|\psi}= S^\dagger_{\psi}S_{\psi}$; the modular Hamiltonian $h_{\psi}\equiv h_{\psi|\psi}=-\log \Delta_{\psi}$.
\end{defn}


\subsection{Algebraic entropies}

In physics literatures, there exists three seemingly different definitions of ``state'':
\begin{enumerate}
	\item A state is a vector $\ket{\psi}$ normalized as $\braket{\psi}{\psi}=1$ in a Hilbert space $\H$.

	\item A state is a density operator (or density matrix) $\rho_{\psi;\A}\in \A$ satisfying 
	\begin{equation}
		\rho_{\psi;\A}=\rho_{\psi;\A}^\dagger;\quad \rho_{\psi;\A}\geq 0;\quad\Tr_{\A}\rho_{\psi;\A}=1.
	\end{equation}

	\item A state is a linear functional $\omega_\psi:\A\rightarrow \mathbb{C}$ satisfying
	\begin{equation}
		\omega_\psi\geq0;\quad \omega_\psi(\id)=1. 
	\end{equation}
\end{enumerate}
These three definitions are correlated to each other via the expectation value: 
\begin{equation}\label{eq:diff def of state}
	\diracprod{\psi}{\O}{\psi}=\Tr_{\A}\left(\rho_{\psi;\A}\O\right)=\omega_\psi(\O),\quad \forall\O\in\A.
\end{equation}
Note that the second definition based on density operators is not well-defined in factors of type III. Also note that, these three definitions are not completely equivalent. In the vector representation, one can always compute the expectation $\diracprod{\psi}{\O}{\psi},~\O\notin\A$ which makes no sense for the other two definitions. To be specific, the expectation value $\Tr_{\A}\left(\rho_{\psi;\A}\O\right),~\O\notin\A$ is ill-defined since $\rho_{\psi;\A}\O$ may not belongs to $\A$ then it is not in the domain of the trace function of $\A$. As for $\omega_\psi(\O),~\O\notin\A$, it is ill-defined since $\omega_\psi$ is only a linear functional on $\A$. Nevertheless, we can extend the last two definitions via \eqref{eq:diff def of state} such that the expectation values of the last two definitions are well-defined for a larger class of operators. We can simply define that
\begin{equation}
	\omega_\psi(\O)=\Tr_{\A}\left(\rho_{\psi;\A}\O\right)= \diracprod{\psi}{\O}{\psi},~\O\notin\A.
\end{equation}
In the following discussions, we always assume the second and the third definitions are extended.

Recall that the quantum relative entropy is defined by
\begin{equation}\label{def:quantum relative entropy}
	\Srel(\rho_\psi|\rho_\phi)\equiv\Tr\left[\rho_{\psi}\left(\log\rho_{\psi}-\log\rho_{\phi}\right)\right]
\end{equation}
which measures how much the state $\psi$ differs from another state $\phi$. Its algebraic generalization is defined due to Araki \cite{araki1975relative,araki1975inequalities}:
\begin{equation}\label{def:Arakis relative entropy}
	\Srel(\psi|\phi;\A)\equiv\diracprod{\psi}{h_{\phi|\psi}}{\psi}=\omega_\psi(h_{\phi|\psi})
\end{equation}
with $\ket{\psi}$ being cyclic and separating with respect to $\A$. See \cite{Witten:2018zxz} for details about how the algebraic relative entropy coincides with the quantum relative entropy in finite-dimensional cases.

The von Neumann entropy is defined by
\begin{equation}\label{eq:vnentropy}
	S(\rho_{\psi})\equiv-\Tr\left(\rho_{\psi}\log\rho_{\psi}\right)
\end{equation}
whose algebraic generalization in factors of type I/II is given by \cite{segal1960note,ohya2004quantum,Longo:2022lod}:
\begin{equation}\label{def:algvnentropy}
	S(\psi;\A)\equiv-\Srel(\psi|\tau;\A)=-\diracprod{\psi}{h_{\tau|\psi}}{\psi}=-\omega_\psi(h_{\tau|\psi})
\end{equation}
where $\ket{\psi}$ is cyclic and separating with respect to $\A$ and $\tau$ is a tracial state. A tracial state is a state $\ket{\tau}\in\H$ satisfying
\begin{equation}
	\diracprod{\tau}{\O\P}{\tau}=\diracprod{\tau}{\P\O}{\tau},\quad \forall\O,\P\in\A
\end{equation}
and each tracial state defines a trace function on $\A$:
\begin{equation}\label{def:trace}
	\Tr_{\A}(\O)\equiv\diracprod{\tau}{\O}{\tau}.
\end{equation}
The trace function is unique up to a rescaling which may contributes to the algebraic von Neumann entropy an additional constant. Such constant does not cause the ambiguity of the algebraic von Neumann entropy, since only the difference between entropies has physical meanings. The non-existence of algebraic von Neumann entropy in factors of type III is due to the non-existence of tracial state or trace function. 

A useful property of the tracial state is
\begin{prop}\label{thm:trace}
	In factors of type I/II, we have a tracial state $\tau$ such that
	\begin{equation}\label{eq:trace}
		\Delta_{\tau|\psi}=\rho_{\psi;\A}^{-1}.
	\end{equation}
	For all states $\psi$.
\end{prop}
We can substitute it into \eqref{def:algvnentropy} and find that the algebraic von Neumann entropy coincides with the usual von Neumann entropy defined in \eqref{eq:vnentropy}.

\section{Page transition via AQEC with complementary recovery}\label{sec:complementary recovery}

Let us give more details about the formulae \eqref{eq:island1} and \eqref{eq:island2}. We first denote the Hilbert space over a Cauchy slice describing an evaporating black in the complete quantum gravity description as $\H^{qg}$, and its semi-classical approximation is denoted as $\H^{sc}$ which excludes quantum gravitational effects, i.e. there exists no gravitational degrees of freedom in $\H^{sc}$. As shown in Fig.\ref{fig:BH}, $\H^{sc}$ is factorized as follows,
\begin{equation}
	\H^{sc}=\H^{sc}_{I\cup R}\otimes \H^{sc}_{X},
\end{equation}
where $\H^{sc}_{I\cup R}\equiv \H^{sc}_{\Sigma_{\text {Rad }} \cup \Sigma_{\text {Island }}}$, $\H^{sc}_{X}\equiv\H^{sc}_{\Sigma_X}$ are the Hilbert spaces of quantum fields on $\Sigma_{\text {Rad }} \cup \Sigma_{\text {Island }},\Sigma_X$ in the semi-classical description, respectively. On the other hand, the complete quantum gravity description also admits a decomposition between the evaporating black hole and its Hawking radiation,
\begin{equation}
	\H^{qg}=\H^{qg}_{\text{Rad}}\otimes\H^{qg}_{\text{BH}}.
\end{equation}

Now consider a state $\tilde{\rho}$ on $\H^{sc}$ as a semi-classical approximation of the true state $\rho$ describing the black hole on $\H^{qg}$, we use $\tilde{\rho}_{I\cup R}$ to denote the reduced density operator of $\tilde{\rho}$ after tracing out the degrees of freedom in $\H^{sc}_{X}$. Similarly, we can define $\tilde{\rho}_{X},\rho_{\text{Rad}}, \rho_{\text{BH}}$. Then the formulae \eqref{eq:island1} and \eqref{eq:island2} can be rewritten as
\begin{equation}\label{eq:island3}
	S(\rho_{\text{Rad}})=\text{min}_X\left\{\text{ext}_X\left[\frac{A_X}{4G}+S(\tilde{\rho}_{I\cup R})\right]\right\},\quad S(\rho_{\text{BH}})=\text{min}_X\left\{\text{ext}_X\left[\frac{A_X}{4G}+S(\tilde{\rho}_{X})\right]\right\},
\end{equation}
where $S(\cdot)$ is the von Neumann entropy evaluated in the respective Hilbert space. One can compare the formulae \eqref{eq:island3} with the statement 4 in the theorem \ref{thm:RecThm2} and see that they share the same form when the solution with non-vanishing island is dominant, i.e. after the Page time, which instructs us to explore whether other statements in the theorem \ref{thm:RecThm2} have their counterparts in the case of black hole evaporation. First of all, we can naturally regard $\H^{qg}=\H^{qg}_{\text{Rad}}\otimes\H^{qg}_{\text{BH}}$ as the physical space and $\H^{sc}=\H^{sc}_{I\cup R}\otimes \H^{sc}_{X}$ as the code space. What follows is that there exists an isometry $V:\H^{sc}\rightarrow \H^{qg}$ which induces two quantum channels:
\begin{equation}\label{eq:channel}
	\begin{aligned}
		&\N: \tilde{\rho}_{I\cup R}\mapsto \rho_{\text{Rad}}\equiv \Tr_{\text{BH}}\left(V \tilde{\rho}_{I\cup R} V^\dagger\right),\\
		&\N': \tilde{\rho}_{X}\mapsto \rho_{\text{BH}}\equiv \Tr_{\text{Rad}}\left(V \tilde{\rho}_{X} V^\dagger\right),
	\end{aligned}
\end{equation}
then for the theorem \ref{thm:RecThm2} to be applicable, we need to argue that
\begin{equation}\label{eq:cond}
	\dim \H^{sc}_{I\cup R}\leq \dim \H^{qg}_{\text{Rad}},\quad \dim \H^{sc}_{X}\leq \dim \H^{qg}_{\text{BH}}
\end{equation}
should hold simultaneously. The first condition in \eqref{eq:cond} is already verified in \cite{Zhong:2024fmn}. The main argument is that, the island region and the radiation region are maximally entangled such that $\dim \H^{sc}_{I\cup R}$ should coincide with the dimension of Hawking radiation in the semi-classical description, which is less than the dimension of Hawking radiation 
in the complete quantum gravity description $\dim\H^{qg}_{\text{Rad}}$ since the former excludes gravitational degrees of freedom. As for the second condition in \eqref{eq:cond}, we assume the central dogma \cite{Almheiri:2020cfm,Strominger:1996sh,Dabholkar:2014ema}:
\begin{itemize}
	\item \emph{for an external observer, a black hole can be treated as an ordinary quantum system with a finite number of degrees of freedom $A_h/(4G)$ where $A_h$ is the area of black hole horizon, and its evolution is unitary.}
\end{itemize}
Since the island boundary $X$ is approximate to the black hole horizon, the central dogma implies that their areas are at the same order such that
\begin{equation}\label{eq:ver1}
	\dim \H^{qg}_{\text{BH}}\sim O\left(\frac{A_X}{4 G}\right),
\end{equation}
which is expected to be large due to the gravitational effect of $O(G^{-1})$. On the other hand, the semi-classical degrees of freedom on $\Sigma_X$ should be sub-leading at the order $O(G^{0})$, so as the von Neumann entropy on $\Sigma_X$, i.e.
\begin{equation}\label{eq:ver2}
	\dim \H^{sc}_{X}\sim O\left(S_{\text {semi-cl }}\left(\Sigma_X\right)\right),\quad \frac{A_X}{4 G}\gg S_{\text {semi-cl }}\left(\Sigma_X\right),
\end{equation}
then \eqref{eq:ver1} together with \eqref{eq:ver2} completes our verification for the condition \eqref{eq:cond} to be held in black hole evaporation.

Now we see that the theorem \ref{thm:RecThm2} can be applied to the case of black hole evaporation after the Page transition,
which further implies by the statement 2 that we have equalities between relative entropies: 
\begin{equation}\label{eq:Sreleq}
\text{(After the Page time)}\quad \Srel(\tilde{\rho}_{I\cup R}|\tilde{\sigma}_{I\cup R})=\Srel(\rho_{\text{Rad}}|\sigma_{\text{Rad}}),~ \Srel(\tilde{\rho}_{X}|\tilde{\sigma}_{X})=\Srel(\rho_{\text{BH}}|\sigma_{\text{BH}})
\end{equation}
for all possible pairs of states $\rho,\sigma$. The equalities \eqref{eq:Sreleq} is equivalent to the fact that the channels \eqref{eq:channel} are exactly reversible due to the theorem \ref{thm:relentropymono}. However, we do not expect \eqref{eq:Sreleq} to hold before the Page transition since during which the solution with vanishing island is dominant. Therefore, the Page transition should correspond to a transition of the exactness for channel recovery. Equivalently we use the theorem \ref{thm:relentropymono} to have
\begin{equation}\label{eq:Srelineq}
	\begin{aligned}
		\Srel(\tilde{\rho}_{I\cup R}(t)|\tilde{\sigma}_{I\cup R}(t))-\Srel(\rho_{\text{Rad}}(t)|\sigma_{\text{Rad}}(t))&\begin{cases}
			>\delta,  & \text{if $t<t_{P}$} \\
			=\delta, & \text{if $t=t_{P}$}\\
			\leq\delta, & \text{if $t>t_{P}$}
		\end{cases},\\
		\Srel(\tilde{\rho}_{X}(t)|\tilde{\sigma}_{X}(t))-\Srel(\rho_{\text{BH}}(t)|\sigma_{\text{BH}}(t))&\begin{cases}
			>\epsilon,  & \text{if $t<t_{P}$} \\
			=\epsilon, & \text{if $t=t_{P}$}\\
			\leq\epsilon, & \text{if $t>t_{P}$}
		\end{cases}.
	\end{aligned}
\end{equation}
for some infinitesimal values $\delta,\epsilon$ as threshold values in approximate quantum error correction. These values are related to the properties of the black hole, and the author in \cite{Zhong:2024fmn} has given an example of calculating these threshold values. Since they are not essential to our present discussions, we will simply assume that they vanish hereafter.

\begin{figure}[h]
	\centering

	\tikzset{every picture/.style={line width=0.75pt}} 
	
	\begin{tikzpicture}[x=0.75pt,y=0.75pt,yscale=-1,xscale=1]
		
		\draw    (150,270) -- (508,270) ;
		\draw [shift={(510,270)}, rotate = 180] [color={rgb, 255:red, 0; green, 0; blue, 0 }  ][line width=0.75]    (10.93,-3.29) .. controls (6.95,-1.4) and (3.31,-0.3) .. (0,0) .. controls (3.31,0.3) and (6.95,1.4) .. (10.93,3.29)   ;
		\draw    (150,270) -- (150,72) ;
		\draw [shift={(150,70)}, rotate = 90] [color={rgb, 255:red, 0; green, 0; blue, 0 }  ][line width=0.75]    (10.93,-3.29) .. controls (6.95,-1.4) and (3.31,-0.3) .. (0,0) .. controls (3.31,0.3) and (6.95,1.4) .. (10.93,3.29)   ;
		\draw [color={rgb, 255:red, 74; green, 144; blue, 226 }  ,draw opacity=1 ]   (470,90) -- (150,270) ;
		\draw [color={rgb, 255:red, 245; green, 166; blue, 35 }  ,draw opacity=1 ]   (150,90) -- (470,270) ;
		\draw [color={rgb, 255:red, 208; green, 2; blue, 27 }  ,draw opacity=1 ]   (160,270) -- (310,190) ;
		\draw [color={rgb, 255:red, 208; green, 2; blue, 27 }  ,draw opacity=1 ]   (450,270) -- (310,190) ;
		\draw  [dash pattern={on 4.5pt off 4.5pt}]  (310,170) -- (310,270) ;
		\draw [color={rgb, 255:red, 65; green, 117; blue, 5 }  ,draw opacity=1 ]   (150,90) .. controls (158.2,214.6) and (243.2,267.6) .. (310,270) ;
		\draw    (470,270) -- (310,270) ;
		\draw  [dash pattern={on 4.5pt off 4.5pt}]  (200,110) -- (200,270) ;
		\draw  [dash pattern={on 4.5pt off 4.5pt}]  (210,270) -- (210,110) ;
		
		\draw (527,272.4) node [anchor=north west][inner sep=0.75pt]    {$t$};
		\draw (307,272.4) node [anchor=north west][inner sep=0.75pt]    {$t_{P}$};
		\draw (241,112.4) node [anchor=north west][inner sep=0.75pt]    {$S_{\text{Rad}}^{\text{island}}$};
		\draw (350,112.4) node [anchor=north west][inner sep=0.75pt]    {$S_{\text{Rad}}^{\text{no-island}}$};
		\draw (369,242.4) node [anchor=north west][inner sep=0.75pt]    {$S_{\text{Rad}}$};
		\draw (156,202.4) node [anchor=north west][inner sep=0.75pt]    {$\Delta S$};
		\draw (191,272.4) node [anchor=north west][inner sep=0.75pt]    {$t_{1}$};
		\draw (212,273.4) node [anchor=north west][inner sep=0.75pt]    {$t_{2}$};

	\end{tikzpicture}
	
	\caption{$\Delta S$ (green line) is defined as the difference between $S_{\text{Rad}}^{\text{island}}$ and $S_{\text{Rad}}$, which monotonically decreases until the Page time $t_P$, i.e. we have $\Delta S(t_2)\leq\Delta S(t_1)$ for $t_1\leq t_2\leq t_P$.}
	\label{fig:Page2}
\end{figure}
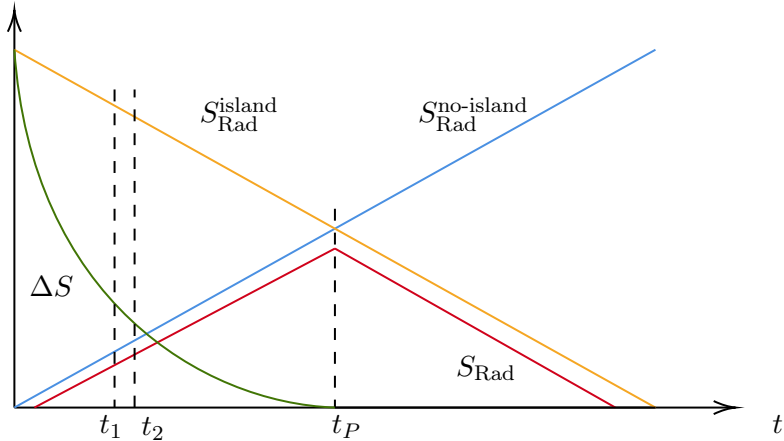

The above arguments by simply applying the reconstruction theorem \ref{thm:RecThm2} may not be convincing for some readers, so we take another detour to directly prove that \eqref{eq:Srelineq} holds for the black hole evaporation. We start with the argument that the entropy of radiation is a competition between two saddle-point solutions,
\begin{equation}
	S(\rho_{\text{Rad}})=\min\left\{S_{\text{Rad}}^{\text{island}},S_{\text{Rad}}^{\text{no-island}}\right\},
\end{equation}
where
\begin{equation}
	S_{\text{Rad}}^{\text{island}}=\frac{A_X}{4G}+S(\tilde{\rho}_{I\cup R}),
\end{equation}
for a non-vanishing island solution $X$. We define
\begin{equation}
	\Delta S_1(\tilde{\rho}(t))\equiv S_{\text{Rad}}^{\text{island}}(t)-S(\rho_{\text{Rad}}(t)),
\end{equation}
which is non-vanishing and monotonically decreasing before $t_{P}$, then vanishes after $t_{P}$:
\begin{equation}\label{eq:DelSprop}
	\Delta S_1(\tilde{\rho}(t))\begin{cases}
		>0,  & \text{if $t<t_{P}$} \\
		=0, & \text{if $t\geq t_{P}$}
	\end{cases},\quad \frac{d\Delta S_1(\tilde{\rho}(t))}{dt}\begin{cases}
		<0,  & \text{if $t<t_{P}$} \\
		=0, & \text{if $t\geq t_{P}$}
	\end{cases},
\end{equation}
see also Fig.\ref{fig:Page2}. Our next step is to rewrite the difference between relative entropies in \eqref{eq:Srelineq} in terms of $\Delta S_1(\rho(t))$, which is already done in \cite{Zhong:2024fmn} and gives the following result,


\begin{equation}
		\Srel(\tilde{\rho}_{I\cup R}(t)|\tilde{\sigma}_{I\cup R}(t))-\Srel(\rho_{\text{Rad}}(t)|\sigma_{\text{Rad}}(t))=\int_{0}^{\tilde{\rho}}d\Delta_1 S(\tilde{\sigma})-\Delta_1 S(\tilde{\rho})
\end{equation}
where $d\Delta_1 S(\tilde{\sigma})\equiv \Delta_1 S(\tilde{\sigma}+d\tilde{\sigma})-\Delta_1 S(\tilde{\sigma})$ and the first term is integrated for $d\tilde{\sigma}$. The positivity of the RHS is guaranteed by the monotonicity in the theorem \ref{thm:relentropymono}. And due to the properties of \eqref{eq:DelSprop}, the RHS monotonically decreases and then vanishes after $t_{P}$, which gives the proof of one half of \eqref{eq:Srelineq}. Similarly, one can define
\begin{equation}
	\Delta S_2(\tilde{\rho}(t))\equiv S_{\text{BH}}^{\text{island}}(t)-S(\rho_{\text{X}}(t)),
\end{equation}
and then follow the same pattern to complete the proof of the other half of \eqref{eq:Srelineq}.

\section{An algebraic description in factors of type I/II}\label{sec:algebraic description}

Algebraically, we consider von Neumann factors of type I/II describing an evaporating black hole, and we introduce $\AIR,\AX,\ARAD,\ABH$ to act on $\HIR,\HX,\HRAD,\HBH$ respectively. We can further assume
\begin{equation}
	\begin{aligned}
		&\AIR=\B(\HIR)\otimes\id_{X},\quad\AX=\id_{I\cup R}\otimes\B(\HX),\\
		& \ARAD=\B(\HRAD)\otimes\id_{\text{BH}},\quad \ABH=\id_{\text{Rad}}\otimes \B(\HBH),
	\end{aligned}
\end{equation} 
such that
\begin{equation}\label{eq:commutant}
	\AIR'=\AX,\quad \ARAD'=\ABH,
\end{equation}
and the algebraic generalization of \eqref{eq:Srelineq} is given by
\begin{equation}\label{eq:algSrelineq}
	\begin{aligned}
		\Srel(\TPsi|\TPhi;\AIR)-\Srel(\Psi|\Phi;\ARAD)&\begin{cases}
			>\delta,  & \text{if $t<t_{P}$} \\
			=\delta, & \text{if $t=t_{P}$}\\
			\leq\delta, & \text{if $t>t_{P}$}
		\end{cases},\\
		\Srel(\TPsi|\TPhi;\AX)-\Srel(\Psi|\Phi;\ABH)&\begin{cases}
			>\epsilon,  & \text{if $t<t_{P}$} \\
			=\epsilon, & \text{if $t=t_{P}$}\\
			\leq\epsilon, & \text{if $t>t_{P}$}
		\end{cases}
	\end{aligned}
\end{equation}
for all states $\TPsi,\TPhi$ on $\H^{sc}$ and $\Psi\equiv V\TPsi,\Phi\equiv V\TPhi$ on $\H^{qg}$, where the relative entropy is promoted to be the algebraic relative entropy. We omit the time-dependence of the states for simplicity, e.g. we write $\TPsi$ to refer to $\TPsi(t)$. The algebraic version \eqref{eq:algSrelineq} of the probe for the quantum channels to be exactly reversible can be similarly argued to hold when applying the algebraic version of the reconstruction theorem \cite{Kang:2018xqy,Xu:2024xbz}. Instead, here we will give explicit calculations as in the end of the last section. First, we promote the von Neumann entropy in \eqref{eq:island3} to be the algebraic von Neumann entropy, such that
\begin{equation}
	S(\Psi;\ARAD)=\min\left\{S_{\text{Rad}}^{\text{island}},S_{\text{Rad}}^{\text{no-island}}\right\},
\end{equation}
where after the Page time we have\footnote{Naively, we expect the algebraic version of the non-vanishing island solution is
\begin{equation}
		S_{\text{Rad}}^{\text{island}}=\frac{A_X}{4G}+S(\TPsi;\AIR).
\end{equation}
However, the fact that the algebraic von Neumann entropy is well-defined in factors of type I/II is equivalent to saying the entropy does not diverge, while the area term on the RHS usually contributes the leading divergence. It is then expected that the algebraic von Neumann entropy should correspond to the usual von Neumann entropy after a subtraction of the area term, i.e. $S(\Psi;\ARAD)\sim S(\rho_{\text{Rad}})-\frac{A_X}{4G}$ for the non-vanishing island solution. See also discussions in \cite{Jensen:2023yxy,Zhong:2024fmn}.
}
\begin{equation}
	S(\Psi;\ARAD)=S_{\text{Rad}}^{\text{island}}\equiv S(\TPsi;\AIR),
\end{equation}
for a non-vanishing island solution $X$. We define
\begin{equation}
	\Delta S_1(\Psi(t))\equiv S_{\text{Rad}}^{\text{island}}(t)-S(\Psi(t);\ARAD),
\end{equation}
which is non-vanishing and monotonically decreasing before $t_{P}$, then vanishes after $t_{P}$: 
\begin{equation}\label{eq:DelSprop2}
	\Delta S_1(\Psi(t))\begin{cases}
		>0,  & \text{if $t<t_{P}$} \\
		=0, & \text{if $t\geq t_{P}$}
	\end{cases},\quad \frac{d\Delta S_1(\Psi(t))}{dt}\begin{cases}
		<0,  & \text{if $t<t_{P}$} \\
		=0, & \text{if $t\geq t_{P}$}
	\end{cases}.
\end{equation}
Therefore, 
\begin{equation}\label{eq:Delta S1}
	S(\Psi;\ARAD)=S(\TPsi;\AIR)-\Delta S_1(\Psi).
\end{equation}
Similarly, we define $\Delta S_2(\Psi(t))$ by
\begin{equation}\label{eq:Delta S2}
	S(\Psi;\ABH)=S(\TPsi;\AX)-\Delta S_2(\Psi),
\end{equation}
which shares the same property as in \eqref{eq:DelSprop2}. Recall that the splittable condition \ref{thm:split} and the property of the trace in \eqref{eq:trace} give
\begin{equation}
	\begin{aligned}
		&\Delta_{\widetilde{\Phi}|\widetilde{\Psi}}=\rho_{\widetilde{\Phi};\AIR}\otimes\rho^{-1}_{\widetilde{\Psi};\AIR'}=\Delta^{-1}_{\widetilde{\tau}|\widetilde{\Phi}}\otimes \Delta_{\widetilde{\tau}|\widetilde{\Psi}}'\\
		\Rightarrow\quad &h_{\widetilde{\Phi}|\widetilde{\Psi}}=-\log\Delta_{\widetilde{\Phi}|\widetilde{\Psi}}=h_{\widetilde{\tau}|\widetilde{\Psi}}'-h_{\widetilde{\tau}|\widetilde{\Phi}},
	\end{aligned}
\end{equation} 
where we define the tracial state $\widetilde{\tau}$ on $\H^{sc}$, and $\Delta_{\widetilde{\Phi}|\widetilde{\Psi}},h_{\widetilde{\Phi}|\widetilde{\Psi}}$ are the relative modular operator and the relative modular Hamiltonian on $\AIR$ respectively. Note that $\Delta_{\widetilde{\tau}|\widetilde{\Psi}}'$ and $h_{\widetilde{\tau}|\widetilde{\Psi}}'$ are the relative modular operator and the relative modular Hamiltonian on $\AX$ respectively, due to \eqref{eq:commutant}. Similarly, we have
\begin{equation}
	h_{{\Phi}|{\Psi}}=h_{{\tau}|{\Psi}}'-h_{{\tau}|{\Phi}},
\end{equation}
where $\tau$ is the tracial state on $\H^{qg}$, and $h_{{\Phi}|{\Psi}},h_{{\tau}|{\Psi}}'$ are the relative modular Hamiltonians on $\ARAD,\ABH$ respectively. Therefore, according to the definitions \eqref{def:Arakis relative entropy} and \eqref{def:algvnentropy} we have
\begin{equation}\label{eq:cal}
	\begin{aligned}
		&\Srel(\TPsi|\TPhi;\AIR)-\Srel(\Psi|\Phi;\ARAD)\\
		=&\diracprod{\TPsi}{h_{\widetilde{\Phi}|\widetilde{\Psi}}}{\TPsi}-\diracprod{\Psi}{h_{{\Phi}|{\Psi}}}{\Psi}\\
		=&S(\Psi;\ABH)-S(\TPsi;\AX)+\diracprod{\Psi}{h_{\tau|\Phi}}{\Psi}-\diracprod{\TPsi}{h_{\widetilde{\tau}|\TPhi}}{\TPsi}\\
		=&\diracprod{\TPsi}{\left( V^\dagger h_{\tau|\Phi} V-h_{\widetilde{\tau}|\TPhi}\right)}{\TPsi}-\Delta S_2(\Psi)\\
		=&\omega_{\widetilde{\Psi}}\left( V^\dagger h_{\tau|\Phi} V-h_{\widetilde{\tau}|\TPhi}\right)-\Delta S_2(\Psi)
	\end{aligned}
\end{equation}
where in the third equality we use $\Psi\equiv V\TPsi$ and \eqref{eq:Delta S2}. Next, we consider a variation of state $\TPsi$ along the direction of $\TPhi$ with an infinitesimal value $\varepsilon$:
\begin{equation}
	\omega_{\widetilde{\Phi}'}=\omega_{\widetilde{\Phi}}+\varepsilon(\omega_{\widetilde{\Psi}}-\omega_{\widetilde{\Phi}}).
\end{equation}
We can use such variation to represent a short period of the black hole evaporation, and we write $\widetilde{\Phi},\widetilde{\Phi}'$ as the states at time $t_1,t_2$ respectively, where $t_1$ and $t_2$ are expected to be close, see Fig.\ref{fig:Page2}. Then the algebraic entropy difference for $\AIR$ is given by
\begin{equation}\label{eq:alg entropy diff1}
	\begin{aligned}
		S(\widetilde{\Phi}';\AIR)-S(\widetilde{\Phi};\AIR)&=\varepsilon(\omega_{\widetilde{\Psi}}-\omega_{\widetilde{\Phi}})(h_{\widetilde{\tau}|\widetilde{\Phi}})\\
		&=\varepsilon~\omega_{\widetilde{\Psi}}(h_{\widetilde{\tau}|\widetilde{\Phi}})+\varepsilon S(\widetilde{\Phi};\AIR).
	\end{aligned}
\end{equation}
On the other hand, we find that $\omega_{\Phi}\equiv\omega_{V\widetilde{\Phi}}$ implies for any operator $\O$,
\begin{equation}
	\begin{aligned}
		\omega_{\Phi'}(\O)&=\omega_{\widetilde{\Phi}'}(V^\dagger\O V)\\
		&=\left[\omega_{\widetilde{\Phi}}+\varepsilon(\omega_{\widetilde{\Psi}}-\omega_{\widetilde{\Phi}})\right](V^\dagger\O V)\\
		&=\omega_{\widetilde{\Phi}}(V^\dagger\O V)+\varepsilon(\omega_{\widetilde{\Psi}}(V^\dagger\O V)-\omega_{\widetilde{\Phi}}(V^\dagger\O V))\\
		&=\omega_{\Phi}(\O )+\varepsilon(\omega_{\Psi}(\O )-\omega_{\Phi}(\O ))\\
		&=\left[\omega_{\Phi}+\varepsilon(\omega_{\Psi}-\omega_{\Phi})\right](\O )
	\end{aligned}
\end{equation}
i.e. we have
\begin{equation}
	\omega_{\Phi'}=\omega_{\Phi}+\varepsilon(\omega_{\Psi}-\omega_{\Phi})
\end{equation}
such that the algebraic entropy difference for $\ARAD$ is given by
\begin{equation}\label{eq:alg entropy diff2}
	\begin{aligned}
		S(\Phi';\ARAD)-S(\Phi;\ARAD)
		&=\varepsilon(\omega_{\Psi}-\omega_{\Phi})(h_{\tau|\Phi})\\
		&=\varepsilon~\omega_{\Psi}(h_{\tau|\Phi})+\varepsilon S(\Phi;\ARAD).
	\end{aligned}
\end{equation}
Note we also have \eqref{eq:Delta S1} such that
\begin{equation}
	S(\Phi';\ARAD)-S(\Phi;\ARAD)=S(\widetilde{\Phi}';\AIR)-S(\widetilde{\Phi};\AIR)+\Delta S_1(\Phi(t_1))-\Delta S_1(\Phi(t_2)).
\end{equation}
Combining it with \eqref{eq:alg entropy diff1} and \eqref{eq:alg entropy diff2}, we have
\begin{equation}
	\varepsilon~\omega_{\Psi}(h_{\tau|\Phi})+\varepsilon S(\Phi;\ARAD)=\varepsilon~\omega_{\widetilde{\Psi}}(h_{\widetilde{\tau}|\widetilde{\Phi}})+\varepsilon S(\widetilde{\Phi};\AIR)+\Delta S_1(\Phi(t_1))-\Delta S_1(\Phi(t_2)),
\end{equation}
which further implies 
\begin{equation}
	\begin{aligned}
		\Rightarrow ~ \varepsilon~\omega_{\widetilde{\Psi}}\left( V^\dagger h_{\tau|\Phi} V-h_{\widetilde{\tau}|\TPhi}\right)&=\varepsilon\left(S(\widetilde{\Phi};\AIR)-S(\Phi;\ARAD)\right)+\Delta S_1(\Phi(t_1))-\Delta S_1(\Phi(t_2))\\
		&=\varepsilon \Delta S_1(\Phi(t_1))+\Delta S_1(\Phi(t_1))-\Delta S_1(\Phi(t_2))
	\end{aligned}
\end{equation}
where we use \eqref{eq:Delta S1} in the second equality. Finally, we continue \eqref{eq:cal} and arrive at
\begin{equation}
	\begin{aligned}
		&\Srel(\TPsi|\TPhi;\AIR)-\Srel(\Psi|\Phi;\ARAD)\\
		&=\omega_{\widetilde{\Psi}}\left( V^\dagger h_{\tau|\Phi} V-h_{\widetilde{\tau}|\TPhi}\right)-\Delta S_2(\Psi(t_1))\\
		&=\frac{1}{\varepsilon}\left[\Delta S_1(\Phi(t_1))-\Delta S_1(\Phi(t_2))\right]+\Delta S_1(\Phi(t_1))-\Delta S_2(\Psi(t_1)).
	\end{aligned}
\end{equation}

As expected, the result is a sum of different $\Delta S_{1,2}$ with the property \eqref{eq:DelSprop2} that they are non-vanishing and monotonically decreasing before $t_{P}$, then vanishes after $t_{P}$. Monotonicity ensures that the term with coefficient $1/\varepsilon$ is non-negative (see also Fig.\ref{fig:Page2}), and it is comparably greater than the other terms due to $\varepsilon\rightarrow0$. Therefore, we conclude that $\Srel(\TPsi|\TPhi;\AIR)-\Srel(\Psi|\Phi;\ARAD)$ satisfies \eqref{eq:Srelineq}. Similar calculations can be performed to show that the other half of \eqref{eq:Srelineq} also holds, which completes our arguments.

\section{Discussions}\label{sec:discussions}

In the paper, we have argued an algebraic generalization of a probe for the Page transition for von Neumann factors of type I/II:
\begin{equation}\label{eq:probe2}
	\begin{aligned}
		\Srel(\TPsi|\TPhi;\AIR)-\Srel(\Psi|\Phi;\ARAD)&\begin{cases}
			>\delta,  & \text{if $t<t_{P}$} \\
			=\delta, & \text{if $t=t_{P}$}\\
			\leq\delta, & \text{if $t>t_{P}$}
		\end{cases},\\
		\Srel(\TPsi|\TPhi;\AX)-\Srel(\Psi|\Phi;\ABH)&\begin{cases}
			>\epsilon,  & \text{if $t<t_{P}$} \\
			=\epsilon, & \text{if $t=t_{P}$}\\
			\leq\epsilon, & \text{if $t>t_{P}$}
		\end{cases}
	\end{aligned}
\end{equation}
which can also be regarded as an algebraic generalization of the island formula describing the evaporating black hole. Similar ideas are also demonstrated in \cite{Gomez:2024fij,Gomez:2024cjm}, where the author have introduced a quantity called Murray von Neumann parameter to describe the algebraic version of the Page curve for type II black holes. The author has shown that, the Page curve describes a phase transition with the transfer of information as order parameter, which coincides with the fact that \eqref{eq:probe2} describes a phase transition of the exactness for channel recovery between different regions of the black hole.

Note that for factors of type III, these are some issues regarding the definitions in describing the Page transition. Especially, the island formulae \eqref{eq:island1} and \eqref{eq:island2} are no longer valid due to the ill-defined von Neumann entropy in factors of type III. Technically, one may introduces the crossed product to construct type II factors from the original type III factors, see also discussions in \cite{Gomez:2024fij,Gomez:2024cjm}. On the other hand, one main advantage of \eqref{eq:probe2} is exactly that the algebraic relative entropy is well-defined in a generic factor, such that \eqref{eq:probe2} is expected to hold in describing type III black holes. Although the direct proof of \eqref{eq:probe2} in type III factors is yet to be established, there are hints for the transfer of information in such cases. For example, the authors in \cite{Engelhardt:2023xer} have interpreted the Page transition as the transfer of an emergent type III subalgebra of high complexity operators from the black hole to radiation. We expect to investigate more formal discussions about \eqref{eq:probe2} on these issues in the future.

\section*{}

\acknowledgments

HZ thanks Qiang Wen, Mingshuai Xu, Yiwei Zhong for helpful discussions. HZ is supported by SEU Innovation Capability Enhancement Plan for Doctoral Students (Grant No. CXJH\_SEU 24137). This work is respectfully dedicated to the memory of HZ's grandfather, who passed away peacefully in January of 2026.

\appendix

\section*{}

\bibliographystyle{JHEP}
\bibliography{bib}

\providecommand{\href}[2]{#2}\begingroup\raggedright\begin{thebibliography}{10}

\bibitem{Page:1993wv}
D.N.~Page, \emph{{Information in black hole radiation}},
  \href{https://doi.org/10.1103/PhysRevLett.71.3743}{\emph{Phys. Rev. Lett.}
  {\bfseries 71} (1993) 3743}
  [\href{https://arxiv.org/abs/hep-th/9306083}{{\ttfamily hep-th/9306083}}].

\bibitem{Page:2013dx}
D.N.~Page, \emph{{Time Dependence of Hawking Radiation Entropy}},
  \href{https://doi.org/10.1088/1475-7516/2013/09/028}{\emph{JCAP} {\bfseries
  09} (2013) 028} [\href{https://arxiv.org/abs/1301.4995}{{\ttfamily
  1301.4995}}].

\bibitem{Mathur:2009hf}
S.D.~Mathur, \emph{{The Information paradox: A Pedagogical introduction}},
  \href{https://doi.org/10.1088/0264-9381/26/22/224001}{\emph{Class. Quant.
  Grav.} {\bfseries 26} (2009) 224001}
  [\href{https://arxiv.org/abs/0909.1038}{{\ttfamily 0909.1038}}].

\bibitem{Almheiri:2012rt}
A.~Almheiri, D.~Marolf, J.~Polchinski and J.~Sully, \emph{{Black Holes:
  Complementarity or Firewalls?}},
  \href{https://doi.org/10.1007/JHEP02(2013)062}{\emph{JHEP} {\bfseries 02}
  (2013) 062} [\href{https://arxiv.org/abs/1207.3123}{{\ttfamily 1207.3123}}].

\bibitem{Penington:2019npb}
G.~Penington, \emph{{Entanglement Wedge Reconstruction and the Information
  Paradox}}, \href{https://doi.org/10.1007/JHEP09(2020)002}{\emph{JHEP}
  {\bfseries 09} (2020) 002}
  [\href{https://arxiv.org/abs/1905.08255}{{\ttfamily 1905.08255}}].

\bibitem{Almheiri:2020cfm}
A.~Almheiri, T.~Hartman, J.~Maldacena, E.~Shaghoulian and A.~Tajdini,
  \emph{{The entropy of Hawking radiation}},
  \href{https://doi.org/10.1103/RevModPhys.93.035002}{\emph{Rev. Mod. Phys.}
  {\bfseries 93} (2021) 035002}
  [\href{https://arxiv.org/abs/2006.06872}{{\ttfamily 2006.06872}}].

\bibitem{Hawking:1975vcx}
S.W.~Hawking, \emph{{Particle Creation by Black Holes}},
  \href{https://doi.org/10.1007/BF02345020}{\emph{Commun. Math. Phys.}
  {\bfseries 43} (1975) 199}.

\bibitem{Hawking:1976ra}
S.W.~Hawking, \emph{{Breakdown of Predictability in Gravitational Collapse}},
  \href{https://doi.org/10.1103/PhysRevD.14.2460}{\emph{Phys. Rev. D}
  {\bfseries 14} (1976) 2460}.

\bibitem{Almheiri:2019hni}
A.~Almheiri, R.~Mahajan, J.~Maldacena and Y.~Zhao, \emph{{The Page curve of
  Hawking radiation from semiclassical geometry}},
  \href{https://doi.org/10.1007/JHEP03(2020)149}{\emph{JHEP} {\bfseries 03}
  (2020) 149} [\href{https://arxiv.org/abs/1908.10996}{{\ttfamily
  1908.10996}}].

\bibitem{Penington:2019kki}
G.~Penington, S.H.~Shenker, D.~Stanford and Z.~Yang, \emph{{Replica wormholes
  and the black hole interior}},
  \href{https://doi.org/10.1007/JHEP03(2022)205}{\emph{JHEP} {\bfseries 03}
  (2022) 205} [\href{https://arxiv.org/abs/1911.11977}{{\ttfamily
  1911.11977}}].

\bibitem{Almheiri:2019qdq}
A.~Almheiri, T.~Hartman, J.~Maldacena, E.~Shaghoulian and A.~Tajdini,
  \emph{{Replica Wormholes and the Entropy of Hawking Radiation}},
  \href{https://doi.org/10.1007/JHEP05(2020)013}{\emph{JHEP} {\bfseries 05}
  (2020) 013} [\href{https://arxiv.org/abs/1911.12333}{{\ttfamily
  1911.12333}}].

\bibitem{Almheiri:2019psf}
A.~Almheiri, N.~Engelhardt, D.~Marolf and H.~Maxfield, \emph{{The entropy of
  bulk quantum fields and the entanglement wedge of an evaporating black
  hole}}, \href{https://doi.org/10.1007/JHEP12(2019)063}{\emph{JHEP} {\bfseries
  12} (2019) 063} [\href{https://arxiv.org/abs/1905.08762}{{\ttfamily
  1905.08762}}].

\bibitem{Zhong:2024fmn}
H.~Zhong, \emph{{Probing the Page transition via approximate quantum error
  correction}}, \href{https://doi.org/10.1007/JHEP01(2025)086}{\emph{JHEP}
  {\bfseries 01} (2025) 086}
  [\href{https://arxiv.org/abs/2408.15104}{{\ttfamily 2408.15104}}].

\bibitem{Almheiri:2014lwa}
A.~Almheiri, X.~Dong and D.~Harlow, \emph{{Bulk Locality and Quantum Error
  Correction in AdS/CFT}},
  \href{https://doi.org/10.1007/JHEP04(2015)163}{\emph{JHEP} {\bfseries 04}
  (2015) 163} [\href{https://arxiv.org/abs/1411.7041}{{\ttfamily 1411.7041}}].

\bibitem{Dong:2016eik}
X.~Dong, D.~Harlow and A.C.~Wall, \emph{{Reconstruction of Bulk Operators
  within the Entanglement Wedge in Gauge-Gravity Duality}},
  \href{https://doi.org/10.1103/PhysRevLett.117.021601}{\emph{Phys. Rev. Lett.}
  {\bfseries 117} (2016) 021601}
  [\href{https://arxiv.org/abs/1601.05416}{{\ttfamily 1601.05416}}].

\bibitem{Harlow:2016vwg}
D.~Harlow, \emph{{The Ryu\textendash{}Takayanagi Formula from Quantum Error
  Correction}}, \href{https://doi.org/10.1007/s00220-017-2904-z}{\emph{Commun.
  Math. Phys.} {\bfseries 354} (2017) 865}
  [\href{https://arxiv.org/abs/1607.03901}{{\ttfamily 1607.03901}}].

\bibitem{Kamal:2019skn}
H.~Kamal and G.~Penington, \emph{{The Ryu-Takayanagi Formula from Quantum Error
  Correction: An Algebraic Treatment of the Boundary CFT}},
  \href{https://arxiv.org/abs/1912.02240}{{\ttfamily 1912.02240}}.

\bibitem{Kang:2018xqy}
M.J.~Kang and D.K.~Kolchmeyer, \emph{{Holographic Relative Entropy in
  Infinite-Dimensional Hilbert Spaces}},
  \href{https://doi.org/10.1007/s00220-022-04627-z}{\emph{Commun. Math. Phys.}
  {\bfseries 400} (2023) 1665}
  [\href{https://arxiv.org/abs/1811.05482}{{\ttfamily 1811.05482}}].

\bibitem{Xu:2024xbz}
M.~Xu and H.~Zhong, \emph{{Adding the algebraic Ryu-Takayanagi formula to the
  algebraic reconstruction theorem}},
  \href{https://arxiv.org/abs/2411.06361}{{\ttfamily 2411.06361}}.

\bibitem{Crann:2024gkv}
J.~Crann and M.J.~Kang, \emph{{Algebraic approach to spacetime bulk
  reconstruction}},  \href{https://arxiv.org/abs/2412.00298}{{\ttfamily
  2412.00298}}.

\bibitem{Maldacena:1997re}
J.M.~Maldacena, \emph{{The Large N limit of superconformal field theories and
  supergravity}}, \href{https://doi.org/10.4310/ATMP.1998.v2.n2.a1}{\emph{Adv.
  Theor. Math. Phys.} {\bfseries 2} (1998) 231}
  [\href{https://arxiv.org/abs/hep-th/9711200}{{\ttfamily hep-th/9711200}}].

\bibitem{Witten:1998qj}
E.~Witten, \emph{{Anti-de Sitter space and holography}},
  \href{https://doi.org/10.4310/ATMP.1998.v2.n2.a2}{\emph{Adv. Theor. Math.
  Phys.} {\bfseries 2} (1998) 253}
  [\href{https://arxiv.org/abs/hep-th/9802150}{{\ttfamily hep-th/9802150}}].

\bibitem{Gubser:1998bc}
S.S.~Gubser, I.R.~Klebanov and A.M.~Polyakov, \emph{{Gauge theory correlators
  from noncritical string theory}},
  \href{https://doi.org/10.1016/S0370-2693(98)00377-3}{\emph{Phys. Lett. B}
  {\bfseries 428} (1998) 105}
  [\href{https://arxiv.org/abs/hep-th/9802109}{{\ttfamily hep-th/9802109}}].

\bibitem{Kang:2019dfi}
M.J.~Kang and D.K.~Kolchmeyer, \emph{{Entanglement wedge reconstruction of
  infinite-dimensional von Neumann algebras using tensor networks}},
  \href{https://doi.org/10.1103/PhysRevD.103.126018}{\emph{Phys. Rev. D}
  {\bfseries 103} (2021) 126018}
  [\href{https://arxiv.org/abs/1910.06328}{{\ttfamily 1910.06328}}].

\bibitem{Faulkner:2020hzi}
T.~Faulkner, \emph{{The holographic map as a conditional expectation}},
  \href{https://arxiv.org/abs/2008.04810}{{\ttfamily 2008.04810}}.

\bibitem{Gesteau:2020hoz}
E.~Gesteau and M.J.~Kang, \emph{{The infinite-dimensional HaPPY code:
  entanglement wedge reconstruction and dynamics}},
  \href{https://arxiv.org/abs/2005.05971}{{\ttfamily 2005.05971}}.

\bibitem{Gesteau:2020rtg}
E.~Gesteau and M.J.~Kang, \emph{{Thermal states are vital: Entanglement Wedge
  Reconstruction from Operator-Pushing}},
  \href{https://arxiv.org/abs/2005.07189}{{\ttfamily 2005.07189}}.

\bibitem{Akers:2021fut}
C.~Akers and G.~Penington, \emph{{Quantum minimal surfaces from quantum error
  correction}},
  \href{https://doi.org/10.21468/SciPostPhys.12.5.157}{\emph{SciPost Phys.}
  {\bfseries 12} (2022) 157}
  [\href{https://arxiv.org/abs/2109.14618}{{\ttfamily 2109.14618}}].

\bibitem{Gesteau:2023hbq}
E.~Gesteau, \emph{{Large N von Neumann Algebras and the Renormalization of
  Newton{\textquoteright}s Constant}},
  \href{https://doi.org/10.1007/s00220-024-05192-3}{\emph{Commun. Math. Phys.}
  {\bfseries 406} (2025) 40}
  [\href{https://arxiv.org/abs/2302.01938}{{\ttfamily 2302.01938}}].

\bibitem{Gomez:2024fij}
C.~Gomez, \emph{{The Algebraic Page Curve}},
  \href{https://arxiv.org/abs/2403.09165}{{\ttfamily 2403.09165}}.

\bibitem{Gomez:2024cjm}
C.~Gomez, \emph{{Duality and Black Hole Evaporation}},
  \href{https://arxiv.org/abs/2405.08435}{{\ttfamily 2405.08435}}.

\bibitem{Engelhardt:2023xer}
N.~Engelhardt and H.~Liu, \emph{{Algebraic ER=EPR and complexity transfer}},
  \href{https://doi.org/10.1007/JHEP07(2024)013}{\emph{JHEP} {\bfseries 07}
  (2024) 013} [\href{https://arxiv.org/abs/2311.04281}{{\ttfamily
  2311.04281}}].

\bibitem{vanderHeijden:2024tdk}
J.~van~der Heijden and E.~Verlinde, \emph{{An Operator Algebraic Approach To
  Black Hole Information}},  \href{https://arxiv.org/abs/2408.00071}{{\ttfamily
  2408.00071}}.

\bibitem{ohya2004quantum}
M.~Ohya and D.~Petz, \emph{Quantum entropy and its use}, Springer Science \&
  Business Media (2004).

\bibitem{Strominger:1996sh}
A.~Strominger and C.~Vafa, \emph{{Microscopic origin of the Bekenstein-Hawking
  entropy}}, \href{https://doi.org/10.1016/0370-2693(96)00345-0}{\emph{Phys.
  Lett. B} {\bfseries 379} (1996) 99}
  [\href{https://arxiv.org/abs/hep-th/9601029}{{\ttfamily hep-th/9601029}}].

\bibitem{Dabholkar:2014ema}
A.~Dabholkar, J.~Gomes and S.~Murthy, \emph{{Nonperturbative black hole entropy
  and Kloosterman sums}},
  \href{https://doi.org/10.1007/JHEP03(2015)074}{\emph{JHEP} {\bfseries 03}
  (2015) 074} [\href{https://arxiv.org/abs/1404.0033}{{\ttfamily 1404.0033}}].

\bibitem{Harlow:2018fse}
D.~Harlow, \emph{{TASI Lectures on the Emergence of Bulk Physics in AdS/CFT}},
  \href{https://doi.org/10.22323/1.305.0002}{\emph{PoS} {\bfseries TASI2017}
  (2018) 002} [\href{https://arxiv.org/abs/1802.01040}{{\ttfamily
  1802.01040}}].

\bibitem{Czech:2012bh}
B.~Czech, J.L.~Karczmarek, F.~Nogueira and M.~Van~Raamsdonk, \emph{{The Gravity
  Dual of a Density Matrix}},
  \href{https://doi.org/10.1088/0264-9381/29/15/155009}{\emph{Class. Quant.
  Grav.} {\bfseries 29} (2012) 155009}
  [\href{https://arxiv.org/abs/1204.1330}{{\ttfamily 1204.1330}}].

\bibitem{Hamilton:2006az}
A.~Hamilton, D.N.~Kabat, G.~Lifschytz and D.A.~Lowe, \emph{{Holographic
  representation of local bulk operators}},
  \href{https://doi.org/10.1103/PhysRevD.74.066009}{\emph{Phys. Rev. D}
  {\bfseries 74} (2006) 066009}
  [\href{https://arxiv.org/abs/hep-th/0606141}{{\ttfamily hep-th/0606141}}].

\bibitem{Morrison:2014jha}
I.A.~Morrison, \emph{{Boundary-to-bulk maps for AdS causal wedges and the
  Reeh-Schlieder property in holography}},
  \href{https://doi.org/10.1007/JHEP05(2014)053}{\emph{JHEP} {\bfseries 05}
  (2014) 053} [\href{https://arxiv.org/abs/1403.3426}{{\ttfamily 1403.3426}}].

\bibitem{Bousso:2012sj}
R.~Bousso, S.~Leichenauer and V.~Rosenhaus, \emph{{Light-sheets and AdS/CFT}},
  \href{https://doi.org/10.1103/PhysRevD.86.046009}{\emph{Phys. Rev. D}
  {\bfseries 86} (2012) 046009}
  [\href{https://arxiv.org/abs/1203.6619}{{\ttfamily 1203.6619}}].

\bibitem{Bousso:2012mh}
R.~Bousso, B.~Freivogel, S.~Leichenauer, V.~Rosenhaus and C.~Zukowski,
  \emph{{Null Geodesics, Local CFT Operators and AdS/CFT for Subregions}},
  \href{https://doi.org/10.1103/PhysRevD.88.064057}{\emph{Phys. Rev. D}
  {\bfseries 88} (2013) 064057}
  [\href{https://arxiv.org/abs/1209.4641}{{\ttfamily 1209.4641}}].

\bibitem{Hubeny:2012wa}
V.E.~Hubeny and M.~Rangamani, \emph{{Causal Holographic Information}},
  \href{https://doi.org/10.1007/JHEP06(2012)114}{\emph{JHEP} {\bfseries 06}
  (2012) 114} [\href{https://arxiv.org/abs/1204.1698}{{\ttfamily 1204.1698}}].

\bibitem{Wall:2012uf}
A.C.~Wall, \emph{{Maximin Surfaces, and the Strong Subadditivity of the
  Covariant Holographic Entanglement Entropy}},
  \href{https://doi.org/10.1088/0264-9381/31/22/225007}{\emph{Class. Quant.
  Grav.} {\bfseries 31} (2014) 225007}
  [\href{https://arxiv.org/abs/1211.3494}{{\ttfamily 1211.3494}}].

\bibitem{Headrick:2014cta}
M.~Headrick, V.E.~Hubeny, A.~Lawrence and M.~Rangamani, \emph{{Causality \&
  holographic entanglement entropy}},
  \href{https://doi.org/10.1007/JHEP12(2014)162}{\emph{JHEP} {\bfseries 12}
  (2014) 162} [\href{https://arxiv.org/abs/1408.6300}{{\ttfamily 1408.6300}}].

\bibitem{Jafferis:2015del}
D.L.~Jafferis, A.~Lewkowycz, J.~Maldacena and S.J.~Suh, \emph{{Relative entropy
  equals bulk relative entropy}},
  \href{https://doi.org/10.1007/JHEP06(2016)004}{\emph{JHEP} {\bfseries 06}
  (2016) 004} [\href{https://arxiv.org/abs/1512.06431}{{\ttfamily
  1512.06431}}].

\bibitem{Polchinski:1999yd}
J.~Polchinski, L.~Susskind and N.~Toumbas, \emph{{Negative energy,
  superluminosity and holography}},
  \href{https://doi.org/10.1103/PhysRevD.60.084006}{\emph{Phys. Rev. D}
  {\bfseries 60} (1999) 084006}
  [\href{https://arxiv.org/abs/hep-th/9903228}{{\ttfamily hep-th/9903228}}].

\bibitem{Almheiri_2015}
A.~Almheiri, X.~Dong and D.~Harlow, \emph{Bulk locality and quantum error
  correction in {AdS}/{CFT}},
  \href{https://doi.org/10.1007/jhep04(2015)163}{\emph{Journal of High Energy
  Physics} {\bfseries 2015} (2015) }.

\bibitem{Ryu:2006bv}
S.~Ryu and T.~Takayanagi, \emph{{Holographic derivation of entanglement entropy
  from AdS/CFT}},
  \href{https://doi.org/10.1103/PhysRevLett.96.181602}{\emph{Phys. Rev. Lett.}
  {\bfseries 96} (2006) 181602}
  [\href{https://arxiv.org/abs/hep-th/0603001}{{\ttfamily hep-th/0603001}}].

\bibitem{Hubeny:2007xt}
V.E.~Hubeny, M.~Rangamani and T.~Takayanagi, \emph{{A Covariant holographic
  entanglement entropy proposal}},
  \href{https://doi.org/10.1088/1126-6708/2007/07/062}{\emph{JHEP} {\bfseries
  07} (2007) 062} [\href{https://arxiv.org/abs/0705.0016}{{\ttfamily
  0705.0016}}].

\bibitem{Casini_2011}
H.~Casini, M.~Huerta and R.C.~Myers, \emph{Towards a derivation of holographic
  entanglement entropy},
  \href{https://doi.org/10.1007/jhep05(2011)036}{\emph{Journal of High Energy
  Physics} {\bfseries 2011} (2011) }.

\bibitem{Lewkowycz:2013nqa}
A.~Lewkowycz and J.~Maldacena, \emph{{Generalized gravitational entropy}},
  \href{https://doi.org/10.1007/JHEP08(2013)090}{\emph{JHEP} {\bfseries 08}
  (2013) 090} [\href{https://arxiv.org/abs/1304.4926}{{\ttfamily 1304.4926}}].

\bibitem{Nishioka:2018khk}
T.~Nishioka, \emph{{Entanglement entropy: holography and renormalization
  group}}, \href{https://doi.org/10.1103/RevModPhys.90.035007}{\emph{Rev. Mod.
  Phys.} {\bfseries 90} (2018) 035007}
  [\href{https://arxiv.org/abs/1801.10352}{{\ttfamily 1801.10352}}].

\bibitem{Faulkner:2013ana}
T.~Faulkner, A.~Lewkowycz and J.~Maldacena, \emph{{Quantum corrections to
  holographic entanglement entropy}},
  \href{https://doi.org/10.1007/JHEP11(2013)074}{\emph{JHEP} {\bfseries 11}
  (2013) 074} [\href{https://arxiv.org/abs/1307.2892}{{\ttfamily 1307.2892}}].

\bibitem{Engelhardt:2014gca}
N.~Engelhardt and A.C.~Wall, \emph{{Quantum Extremal Surfaces: Holographic
  Entanglement Entropy beyond the Classical Regime}},
  \href{https://doi.org/10.1007/JHEP01(2015)073}{\emph{JHEP} {\bfseries 01}
  (2015) 073} [\href{https://arxiv.org/abs/1408.3203}{{\ttfamily 1408.3203}}].

\bibitem{Vaughan2009}
V.F.~Jones, ``Von neumann algebras.'' https://math.berkeley.edu/~vfr/MATH20909/
  VonNeumann2009.pdf, 2009.

\bibitem{Sorce:2023fdx}
J.~Sorce, \emph{{Notes on the type classification of von Neumann algebras}},
  \href{https://doi.org/10.1142/S0129055X24300024}{\emph{Rev. Math. Phys.}
  {\bfseries 36} (2024) 2430002}
  [\href{https://arxiv.org/abs/2302.01958}{{\ttfamily 2302.01958}}].

\bibitem{reed1972methods}
M.~Reed, B.~Simon, B.~Simon and B.~Simon, \emph{Methods of modern mathematical
  physics}, vol.~1, Academic press New York (1972).

\bibitem{takesaki2006tomita}
M.~Takesaki, \emph{Tomita's theory of modular Hilbert algebras and its
  applications}, vol.~128, Springer (2006).

\bibitem{Witten:2018zxz}
E.~Witten, \emph{{APS Medal for Exceptional Achievement in Research: Invited
  article on entanglement properties of quantum field theory}},
  \href{https://doi.org/10.1103/RevModPhys.90.045003}{\emph{Rev. Mod. Phys.}
  {\bfseries 90} (2018) 045003}
  [\href{https://arxiv.org/abs/1803.04993}{{\ttfamily 1803.04993}}].

\bibitem{Jensen:2023yxy}
K.~Jensen, J.~Sorce and A.J.~Speranza, \emph{{Generalized entropy for general
  subregions in quantum gravity}},
  \href{https://doi.org/10.1007/JHEP12(2023)020}{\emph{JHEP} {\bfseries 12}
  (2023) 020} [\href{https://arxiv.org/abs/2306.01837}{{\ttfamily
  2306.01837}}].

\bibitem{Faulkner:2024gst}
T.~Faulkner and A.J.~Speranza, \emph{{Gravitational algebras and the
  generalized second law}},
  \href{https://doi.org/10.1007/JHEP11(2024)099}{\emph{JHEP} {\bfseries 11}
  (2024) 099} [\href{https://arxiv.org/abs/2405.00847}{{\ttfamily
  2405.00847}}].

\bibitem{araki1975relative}
H.~Araki, \emph{Relative entropy of states of von neumann algebras},
  {\emph{Publications of the Research Institute for Mathematical Sciences}
  {\bfseries 11} (1975) 809}.

\bibitem{araki1975inequalities}
H.~Araki, \emph{Inequalities in von neumann algebras}, {\emph{Les rencontres
  physiciens-math{\'e}maticiens de Strasbourg-RCP25} {\bfseries 22} (1975) 1}.

\bibitem{segal1960note}
I.E.~Segal, \emph{A note on the concept of entropy}, {\emph{Journal of
  Mathematics and Mechanics} (1960) 623}.

\bibitem{Longo:2022lod}
R.~Longo and E.~Witten, \emph{{A note on continuous entropy}},
  \href{https://doi.org/10.4310/PAMQ.2023.v19.n5.a5}{\emph{Pure Appl. Math.
  Quart.} {\bfseries 19} (2023) 2501}
  [\href{https://arxiv.org/abs/2202.03357}{{\ttfamily 2202.03357}}].

\end{thebibliography}\endgroup
\end{document}